\documentclass[10pt,twocolumn,letterpaper]{article}

\usepackage[pagenumbers]{wacv} 

\newif\ifreview
\reviewfalse

\definecolor{wacvblue}{rgb}{0.21,0.49,0.74}
\usepackage[pagebackref,breaklinks,colorlinks,allcolors=wacvblue]{hyperref}

\def\wacvPaperID{713} 
\def\confName{WACV}
\def\confYear{2026}

\title{Seeing Beyond the Lesion: Disease Recognition from Reactive CNS Tissue}

\author{%
\parbox{\linewidth}{\centering
Jan Schnorrenberg\textsuperscript{1,\dag}\quad
Jan Ernsting\textsuperscript{2,3,4,5\dag}\quad
Enrico Küllenberg\textsuperscript{1}\quad
Tim Hahn\textsuperscript{4,5}\\
Benjamin Risse\textsuperscript{2,3,\ddag}\quad
Christian Thomas\textsuperscript{1,\ddag}\\[4pt]
\textsuperscript{1}Institute of Neuropathology, University Hospital Münster, Münster, Germany\\
\textsuperscript{2}Institute for Geoinformatics, University of Münster, Münster, Germany\\
\textsuperscript{3}{Faculty of Mathematics and Computer Science, University of Münster, Münster, Germany}\\
\textsuperscript{4}{Institute for Machine Learning in Medicine (Focus Area Psychiatry), University of Münster, Münster, Germany}\\
\textsuperscript{5}{University of Münster, Institute for Translational Psychiatry, Münster, Germany}
\\[2pt]
{\tt\small \{jan.schnorrenberg, j.ernsting, hahnt, b.risse\}@uni-muenster.de}\\
{\tt\small \{enrico.kuellenberg, christian.thomas\}@ukmuenster.de}%
}}

\begin{document}
\maketitle
\ifreview\else
\renewcommand{\thefootnote}{}
\footnotetext[1]{\textsuperscript{\dag}These authors contributed equally.}
\footnotetext[2]{\textsuperscript{\ddag}These authors contributed equally as senior authors.}
\renewcommand{\thefootnote}{\arabic{footnote}}
\fi
\begin{abstract}
Sampling error yields exclusively reactive, non-lesional brain parenchyma in a significant proportion of intracranial biopsies, leaving the underlying disease undiagnosed. 
We benchmark four pathology foundation models (UNI2-h, Virchow2, Prov-GigaPath, H-optimus-0) as frozen patch encoders within a shared attention-based multiple-instance learning framework using 245 whole-slide images from 186 patients with confirmed downstream diagnoses.
We first show that coarse disease-category prediction can be reproduced largely from slide size alone.
After restricting classification to three finer diagnostic distinctions within common tissue categories, this confound no longer explains performance, yet disease labels remain predictable above chance under permutation testing (p $\le 10^{-4}$ throughout).
Surprisingly, performance is statistically indistinguishable across all foundation-model encoders, suggesting that recovering these weak morphological signatures is not limited by current patch representations.
Signed instance-contribution maps and expert review further test whether predictive evidence localizes to reactive parenchyma rather than sampling-induced bias like blood introduced during tissue sampling.
These results position acquisition-shortcut auditing via a provenance-only baseline as a necessary control in computational-pathology benchmarks, and show, once that confound is removed, that weakly supervised models still recover disease signal from tissue conventionally regarded as non-diagnostic.
\end{abstract}

\section{Introduction}
\label{sec:intro}

Brain biopsy is essential for diagnosing intracranial lesions, yet sampling error yields only reactive, non-lesional brain parenchyma in up to $14\%$ of procedures~\cite{Pasternak2021-dj}. 
In these cases, conventional histopathological evaluation typically reveals reactive gliosis, including activated astrocytes and microglial proliferation, but cannot identify the underlying disease process responsible for the tissue response. 
Reactive gliosis itself is biologically heterogeneous, with context-dependent astrocytic and microglial states identified by single-cell and spatial transcriptomic studies~\cite{Hasel2021-pa,Keren-Shaul2017-we}. 
These coordinated cellular responses raise the possibility that pathological processes such as primary central nervous system (CNS) tumors, metastases or brain abscesses may induce distinct, albeit subtle, morphological patterns in the surrounding reactive brain tissue.

Deep learning has shown that routine hematoxylin and eosin-stained whole-slide images (WSIs) encode diagnostic and molecular information beyond features readily recognized by human observers. 
Because WSIs are gigapixel-scale and annotations are typically available only at the slide or case level, multiple instance learning (MIL) has become a standard weakly supervised learning framework for computational pathology. 
In MIL, a slide is represented as a collection of image patches whose features are aggregated into a slide-level representation, allowing informative regions to be identified without exhaustive spatial annotation~\cite{Campanella2019-dm,Lu2021-lh}.

This paradigm has recently achieved impressive performance in CNS tumor classification. 
DEPLOY predicted broad DNA methylation-defined CNS tumor classes directly from histology~\cite{Hoang2024-rt}, while Hetairos combined patch embeddings from the Prov-GigaPath foundation model with a transformer-based MIL aggregator to classify 102 methylation-defined CNS tumor subtypes in a large multicenter cohort~\cite{Jin2026-bq}. 
These approaches demonstrate that weakly supervised learning can capture diagnostically relevant morphological information from whole-slide images. 
However, they are designed for overtly lesional tissue, where neoplastic morphology provides a dominant class-defining signal. 
In contrast, reactive-only biopsy specimens present a fundamentally different learning problem: any disease-specific information is expected to be subtle, spatially diffuse, and embedded within tissue that lacks readily apparent pathological hallmarks. 
This setting is particularly well suited to MIL, which can integrate weak evidence across the entire tissue section using only slide-level labels.

The quality of image representations is likely to become increasingly important as the discriminative signal weakens.
Recent pathology foundation models have learned highly transferable visual representations from large and diverse histopathology datasets~\cite{Chen2024-fc,Xu2024-hm,Vorontsov2024-dq}. 
However, these models differ substantially in their pretraining data, learning objectives, architectures, and receptive fields, and their ability to capture subtle reactive CNS morphology remains unknown.
Performance on tumor-centered benchmarks therefore cannot be assumed to generalize to reactive, non-lesional brain tissue.

Here, we approached the differential diagnosis of reactive gliosis as a three-class, weakly supervised WSI classification task. 
Reactive brain tissue was sampled adjacent to the underlying lesion.
We used MIL to distinguish the samples into tumor, inflammation and control, while systematically comparing multiple pathology foundation models as feature encoders within a common slide-level learning framework. 
We further refined this classification to differentiate among primary CNS tumor entities and to infer the tissue of origin of metastatic tumors. 
By focusing exclusively on reactive, non-lesional tissue, this study shows that the underlying disease process can be inferred indirectly from tissue responses alone and evaluates how representation learning influences performance when discriminative morphological signals are subtle and spatially distributed.

Our contributions are threefold: (i) we show that a provenance-only (slide-size) baseline exposes an acquisition shortcut that inflates apparent performance on the coarse task, and argue it belongs in computational-pathology evaluation as a standard control; (ii) once this confound is removed, we establish via permutation testing that disease signal is recoverable above chance from reactive, non-lesional tissue across three finer tasks; and (iii) under corrected significance testing, four state-of-the-art encoders are statistically indistinguishable, indicating representation choice is not the current bottleneck.
\section{Related work}
\label{sec:relatedwork}

\paragraph{Weakly supervised computational pathology.}
Whole-slide images are gigapixel-scale and typically annotated only at the slide or case level, which has motivated multiple instance learning (MIL) as framework for computational pathology~\cite{Campanella2019-dm}. 
Attention-based MIL learns to weight individual patches when forming a slide-level representation~\cite{Ilse2018-wd}, and clustering- and attention-constrained variants such as CLAM extend this to data-efficient, weakly supervised classification while exposing the regions that drive a prediction~\cite{Lu2021-lh}. 
These methods are particularly well suited to reactive, non-lesional tissue, in which disease-associated morphological signals may be subtle, spatially distributed and unavailable for direct annotation.

\paragraph{Pathology foundation models.}
Self-supervised pretraining on large, diverse histopathology datasets has produced general-purpose patch encoders that transfer across many downstream tasks~\cite{Chen2024-fc,Xu2024-hm,Vorontsov2024-dq}.
The models we benchmarked in our study, namely UNI2-h~\cite{Chen2024-lr}, Prov-GigaPath~\cite{Xu2024-tm}, Virchow2~\cite{Zimmermann2024-na}, and H-optimus-0~\cite{hoptimus0}, differ substantially in pretraining data, learning objectives, architectures, and receptive fields.
Their performance has been assessed on tumor-centered benchmarks, in which neoplastic morphology provides a strong discriminative signal. Whether these representations can capture the subtle, spatially distributed morphology of reactive brain tissue remains unclear, and performance on lesional tasks cannot be assumed to generalize to this setting.

\paragraph{Deep learning for CNS tumor histology.}
Weakly supervised models have recently achieved strong performance in CNS tumor classification directly from H\&E histology.
DEPLOY predicts broad DNA methylation-defined tumor classes from whole-slide images~\cite{Hoang2024-rt}, and Hetairos couples Prov-GigaPath patch embeddings with a transformer-based MIL aggregator to resolve a large number of methylation-defined subtypes across a multicenter cohort~\cite{Jin2026-bq}.
These approaches analyze tissue with morphologically evident lesions, in which neoplastic features provide the dominant class-defining signal.
Our setting poses a fundamentally different challenge: we classify reactive, non-lesional brain parenchyma lacking readily recognizable diagnostic hallmarks, such that disease-specific information must be inferred indirectly from the tissue response. 
The biological plausibility of this approach is supported by single-cell and spatial studies showing that reactive astrocytic and microglial states are heterogeneous and context-dependent~\cite{Hasel2021-pa,Keren-Shaul2017-we}.
To our knowledge, the classification of underlying disease processes based exclusively on reactive, non-lesional CNS tissue has not previously been investigated.

\section{Methods}
\label{sec:methods}

\subsection{Datasets}
\begin{table}
    \centering
    \begin{tabular}{lrr}
    \toprule
    \textbf{Classification} & \textbf{Patients} & \textbf{Slides} \\
    \midrule
    \textbf{Tumor}                        & \textbf{128} & \textbf{177} \\
    \quad Diffuse gliomas                 & 59  & 84 \\
    \quad\quad Astrocytoma                & 20  & 24 \\
    \quad\quad Oligodendroglioma          & 19  & 24 \\
    \quad\quad Glioblastoma               & 20  & 36 \\
    \quad Metastases                      & 69  & 93 \\
    \quad\quad Lung                       & 19  & 21 \\
    \quad\quad Colon                      & 11  & 15 \\
    \quad\quad Breast                     & 19  & 24 \\
    \quad\quad Skin                       & 20  & 33 \\
    \textbf{Inflammation}                 & \textbf{28} & \textbf{38} \\
    \textbf{Control}                      & \textbf{30} & \textbf{30} \\
    \midrule
    \textbf{Total}                        & \textbf{186} & \textbf{245} \\
    \bottomrule
    \end{tabular}
    \caption{Hierarchical composition of tissue types in the curated dataset, reported as unique patients and whole-slide images. Each hierarchy level supplies the labels for one classification task: 
    the three top-level types (tumor, inflammation, control) for \emph{Category};
    the two tumor entities (glioma vs.\ metastasis) for \emph{Subtype}; 
    the three glioma entities for \emph{Glioma}; 
    and the four metastatic origins for \emph{Origin} (cf.\ \autoref{tab:results}). 
    The dataset comprises 186 patients and 245 slides.}
    \label{tab:dataset}
\end{table}

H\&E-stained whole-slide images (WSIs) from 245 samples were retrieved from the archive of the Institute of Neuropathology, University Hospital Münster.
The use of anonymized biopsy specimens for research complied with local regulations of the University Hospital Münster and was approved by the Münster ethics committee (2026-346-f-S).

\paragraph{Ground-truth generation for reactive tissue.}
Because this study focuses on reactive, non-lesional brain parenchyma rather than the lesions themselves, each WSI had to be reduced to tissue that carries a disease label without containing the lesional tissue. 
We employed two complementary strategies. 
For most tumor cases, the primary lesion and adjacent reactive tissue were present on the same slide. In these cases, we removed the lesional region using the Aperio ImageScope software (Leica Biosystems), retaining only the adjacent reactive parenchyma. The resulting reactive-tissue WSIs were often small and therefore yielded relatively few image patches.
We additionally identified cases in which an initial biopsy was reported as non-diagnostic (i.e., "reactive gliosis'' or "non-specific inflammation'') while a subsequent biopsy of the same patient, captured on a separate slide, established a specific diagnosis. 
These cases were particularly valuable because they contained tissue that had been considered genuinely non-diagnostic by the reporting neuropathologist, yet could be linked to a definitive ground-truth label. They also generally provided substantially larger tissue areas than the manually edited reactive regions.
Because these two sampling strategies yielded systematically different amounts of tissue, slide size was correlated with sampling provenance and could therefore provide a shortcut feature for classification. 
We therefore quantified and controlled for this potential confound using a size-only baseline.

\subsection{Preprocessing}
WSIs, originally acquired in \texttt{svs} or \texttt{ndpi} format, were converted to BigTIFF using \textit{libvips}~\cite{1530120}.
During conversion, the imaging data was down-sampled from the native $40\times$ magnification ($0.233\,\mu$m/pixel) to $20\times$ magnification ($0.5\,\mu$m/pixel) to correspond to the nominal resolution the foundation models were trained on.

\subsection{Embedders}
We evaluated four pathology foundation models as patch-level feature encoders: UNI2-h~\cite{Chen2024-lr}, Virchow2~\cite{Zimmermann2024-na}, Prov-GigaPath~\cite{Xu2024-tm}, and H-optimus-0~\cite{hoptimus0}. 
UNI2-h, Prov-GigaPath, and H-optimus-0 produce 1536-dimensional patch embeddings, whereas Virchow2 produces 2560-dimensional embeddings.
We treated the embedders as frozen embedding models and did not use any end-to-end training or fine tuning of the embedding model to the specific task.

\subsection{Gated MIL attention}
\label{sec:mil}
We framed each slide as a bag $X = \{x_1, \dots, x_K\}$ of $K$ tissue patches with a single slide-level label.
A frozen pathology foundation model $f$ (\autoref{sec:methods} embedders) maps each patch to an embedding $h_k = f(x_k) \in \mathbb{R}^{d}$, which a trainable layer projects to a lower-dimensional feature $\tilde{h}_k = \mathrm{ReLU}(W_0 h_k) \in \mathbb{R}^{d'}$, with $d' = 256$ identical across embedders, so that only $W_0 \in \mathbb{R}^{d' \times d}$ absorbs the differing embedder dimensions $d$ while the aggregation and classifier were identical across encoders.
Patches were aggregated with the gated-attention mechanism of Ilse et al.~\cite{Ilse2018-wd}, the pooling backbone shared by many attention-based WSI-MIL methods~\cite{Lu2021-dy}. 
Unlike CLAM~\cite{Lu2021-dy}, which augments this backbone with multi-branch attention and an instance-level clustering loss, we retained the plain single-branch classifier.

We retained the plain single-branch classifier for two reasons.
First, its minimality lets the per-patch contribution decomposition (below) hold exactly.
Second, gated attention pools tiles without any spatial or relational inductive bias, so differences in benchmark performance track the encoder's feature space rather than the capacity of the aggregator, which is what a benchmark of frozen foundation-model encoders should surface~\cite{Campanella2025-xh}. 
Gated-attention MIL is, accordingly, a standard aggregator for evaluating pathology foundation models~\cite{Chen2024-au,Vorontsov2024-yi} and stays competitive with more recent designs~\cite{pmlr-v254-chen24a}.

Each patch receives a normalized, non-negative attention weight
\begin{equation}
    a_k = \frac{\exp\!\big\{\, w^{\top}\big(\tanh(V\tilde{h}_k) \odot \sigma(U\tilde{h}_k)\big)\big\}}{\sum_{j=1}^{K}\exp\!\big\{\, w^{\top}\big(\tanh(V\tilde{h}_j) \odot \sigma(U\tilde{h}_j)\big)\big\}},
\end{equation}
where $U, V \in \mathbb{R}^{m \times d'}$ and $w \in \mathbb{R}^{m}$ are learnable, $\sigma$ is the sigmoid gate, and $\odot$ denotes elementwise multiplication; bias terms are omitted.
The slide representation is the attention-weighted mean $z = \sum_{k=1}^{K} a_k \tilde{h}_k$, and a linear classifier produces the class logits $s = W_c z + b_c$, with $s \in \mathbb{R}^{C}$ over the $C$ tissue classes.
The network was trained with a class-balanced cross-entropy loss on the slide label, where each class $c$ is weighted by $\omega_c = N / (C\, n_c)$, with $N$ the number of training slides and $n_c$ the number of training slides of class $c$.

\paragraph{Signed patch contribution scores.}
Inspired by~\cite{javed2022additive}, we derived an interpretable heatmap from per-patch contributions. 
In contrast to their additive head, here the decomposition is already exact for standard gated-attention MIL, since pooling is a weighted sum and the classifier is linear.
The logit for class $c$ therefore splits into per-patch terms:
\begin{equation}
    s_c = \sum_{k=1}^{K} \underbrace{a_k \, w_c^{\top} \tilde{h}_k}_{\phi_{k,c}} \; + \; b_c,
\end{equation}
where $w_c$ is the $c$-th row of $W_c$ and $\phi_{k,c}$ is the contribution of patch $k$ to class $c$.
Unlike the attention weight $a_k$, which is strictly non-negative and therefore only encodes salience, the contribution $\phi_{k,c}$ carries a sign: $\phi_{k,c} > 0$ marks a patch as evidence \emph{for} class $c$, while $\phi_{k,c} < 0$ marks it as evidence \emph{against}. 
Its magnitude $|\phi_{k,c}|$ couples attention salience $a_k$ with the class-aligned projection $w_c^{\top}\tilde{h}_k$.
Mapping $\phi_{k,c}$ back to patch coordinates yields class-specific, signed contribution heatmaps that localize where and in which direction reactive tissue drives each prediction, which is well suited to the subtle, spatially diffuse signal expected in non-lesional brain parenchyma.

\subsection{Training}
All models were trained for $14$ epochs with AdamW and a cosine learning-rate schedule (peak $5\times10^{-4}$, final $10^{-6}$) preceded by $5$ warmup epochs. 
Weight decay follows a cosine schedule from $0.04$ to $0.4$, with bias and normalization parameters excluded.
Because bags contain a variable number of patches, we used a batch size of 1 and accumulated gradients over 8 slides (effective batch size 8), optimizing the class-balanced cross-entropy loss of \autoref{sec:mil}.
We reported the final-epoch model without early stopping, so that no epoch is selected on the validation fold.
Each fold was fully seeded (split, initialization, and data order) for determinism.

All experiments were conducted using Monte Carlo cross-validation.
For each task we drew a fixed number of independent, patient-stratified train/test splits (\texttt{StratifiedShuffleSplit}, 80\% of patients train, 20\% held out), with all slides from a given patient assigned to the same split to prevent leakage
between training and evaluation.
The number of draws was capped by the smallest class at the patient level, or at 20: 20 for category-level and tumor-subtype classification, 19 for glioma-typing, and 11 for metastatic-origin.
For metastatic origin this yields roughly 14 held-out patients ($\approx$19 slides) per draw.
Because the draws are independent, a patient may appear in several held-out splits or in none;
the metastatic-origin evaluation therefore covers the 78 of 93 metastasis slides whose patient was drawn into at least one test fold, with per-class support of 21/13/16/28 slides for breast/colon/lung/skin.

Training was parallelized with GNU Parallel~\cite{Tange2026-yj}, fitting all folds concurrently on a single NVIDIA RTX 4090 GPU.

\subsection{Statistical analysis.}
We assessed significance with two complementary tests, computed directly from the per-fold slide-level predictions.
To test whether each encoder performs \emph{above chance} on a given task, we used a conditional label-permutation test: 
holding the trained model's predictions fixed, we permuted the ground-truth labels within each fold, recomputed the metric, and averaged it across folds to obtain one draw from the null distribution. 
Repeating this $P=10,000$ times yields a null against which the observed mean metric is compared, with $p = (1 + \#\{\text{null} \ge \text{observed}\})/(P+1)$. 
Because the predictions are held fixed, this tests whether the model's outputs are associated with the labels beyond chance, rather than re-estimating training variability.

To compare encoders, we tested each pairwise difference in per-fold metrics with the Nadeau--Bengio corrected resampled $t$-test~\cite{Nadeau2003-eh}, which inflates the naive paired-$t$ variance by $\big(\tfrac{1}{J} + \tfrac{n_{\text{test}}}{n_{\text{train}}}\big)$ to account for the training-set overlap between Monte Carlo cross-validation folds.
A plain paired $t$-test is anticonservative in this setting.
The $J$ folds are paired across encoders by identical test slides, verified per fold. Within each task and metric, the six pairwise $p$-values were Holm-corrected for multiple comparisons.


\subsection{Attention-tile artifact annotation}
\label{sec:blood-annotation}
To assess whether top-ranked attention localized to disease-relevant tissue rather than preparation artifacts, a board-certified neuropathologist (single reader) visually reviewed the five highest-contribution tiles per slide for each encoder and labeled each tile as containing predominantly sampling-introduced blood ($>50\%$ erythrocytes) or not.
For each encoder we report the mean fraction of these top-5 tiles flagged as blood, averaged over the reviewed slides.
The analysis was performed on the category task, which comprises the most slides and therefore affords the greatest statistical power for a per-tile comparison.
Because splits are patient-stratified and several patients contribute multiple slides, review covered the 183 of 245 category slides whose patient was drawn into at least one test fold.
Every slide was reviewed under all four encoders, yielding complete blocks (encoders as related conditions, slides as blocks), so we compared the per-slide blood fractions across encoders with a Friedman test and report the resulting $\chi^2$ statistic (with $k-1$ degrees of freedom for $k=4$ encoders) and $p$-value.
Because annotation used a single reader without a second independent rater, we do not report inter-rater agreement, and the absolute fractions should be read as descriptive; the cross-encoder comparison is intended to characterize relative differences in artifact susceptibility rather than to establish an absolute artifact rate.

\section{Results}
\label{sec:results}

\newcommand{\std}[1]{{\scriptsize$\,\pm#1$}}

\begin{table*}
    \centering
    \small
    \setlength{\tabcolsep}{4pt}
    \begin{tabular}{lcccccccc}
    \toprule
    & \multicolumn{4}{c}{\textbf{Balanced accuracy}} & \multicolumn{4}{c}{\textbf{AUC}} \\
    \cmidrule(lr){2-5}\cmidrule(lr){6-9}
    \textbf{Encoder} & Category & Subtype & Origin$^{\dagger}$ & Glioma$^{\ddagger}$ & Category & Subtype & Origin$^{\dagger}$ & Glioma$^{\ddagger}$ \\
    \midrule
    H-optimus-0    & 0.792\std{0.066} & \textbf{0.819}\std{0.070} & 0.406\std{0.166} & 0.557\std{0.122} & 0.912\std{0.031} & 0.913\std{0.045} & 0.686\std{0.096} & 0.723\std{0.100} \\
    Prov-GigaPath  & 0.797\std{0.068} & 0.809\std{0.070} & 0.394\std{0.117} & 0.552\std{0.158} & \textbf{0.921}\std{0.031} & 0.906\std{0.051} & \textbf{0.697}\std{0.072} & \textbf{0.747}\std{0.109} \\
    UNI2-h         & \textbf{0.798}\std{0.062} & 0.807\std{0.082} & 0.371\std{0.132} & \textbf{0.581}\std{0.114} & 0.919\std{0.033} & 0.911\std{0.049} & 0.652\std{0.074} & 0.726\std{0.114} \\
    Virchow2       & 0.798\std{0.056} & 0.819\std{0.071} & \textbf{0.416}\std{0.118} & 0.437\std{0.131} & 0.916\std{0.032} & \textbf{0.921}\std{0.049} & 0.697\std{0.090} & 0.655\std{0.104} \\
    \midrule
    \textit{Size only} & \textit{0.782}\std{0.042} & \textit{0.518}\std{0.094} & \textit{0.186}\std{0.094} & \textit{0.361}\std{0.098} & \textit{0.877}\std{0.046} & \textit{0.535}\std{0.099} & \textit{0.543}\std{0.063} & \textit{0.481}\std{0.088} \\
    \textit{Chance}    & \textit{0.333} & \textit{0.500} & \textit{0.250} & \textit{0.333} & \textit{0.500} & \textit{0.500} & \textit{0.500} & \textit{0.500} \\
    \bottomrule
    \end{tabular}
    \caption{Slide-level classification across the four hierarchical tasks (mean\,$\pm$\,std over Monte Carlo cross-validation folds at the final training epoch; best encoder per task in \textbf{bold}). The \emph{size-only} row is a logistic-regression baseline trained on $\log$ tile count alone, using identical folds and metrics; it quantifies how much of each task is recoverable from slide size without tissue morphology. $^{\dagger}$Origin is evaluated over 11 folds and $^{\ddagger}$glioma over 19; category and subtype over 20.}
    \label{tab:results}
\end{table*}

\begin{figure*}
    \centering
    \includegraphics[width=\linewidth]{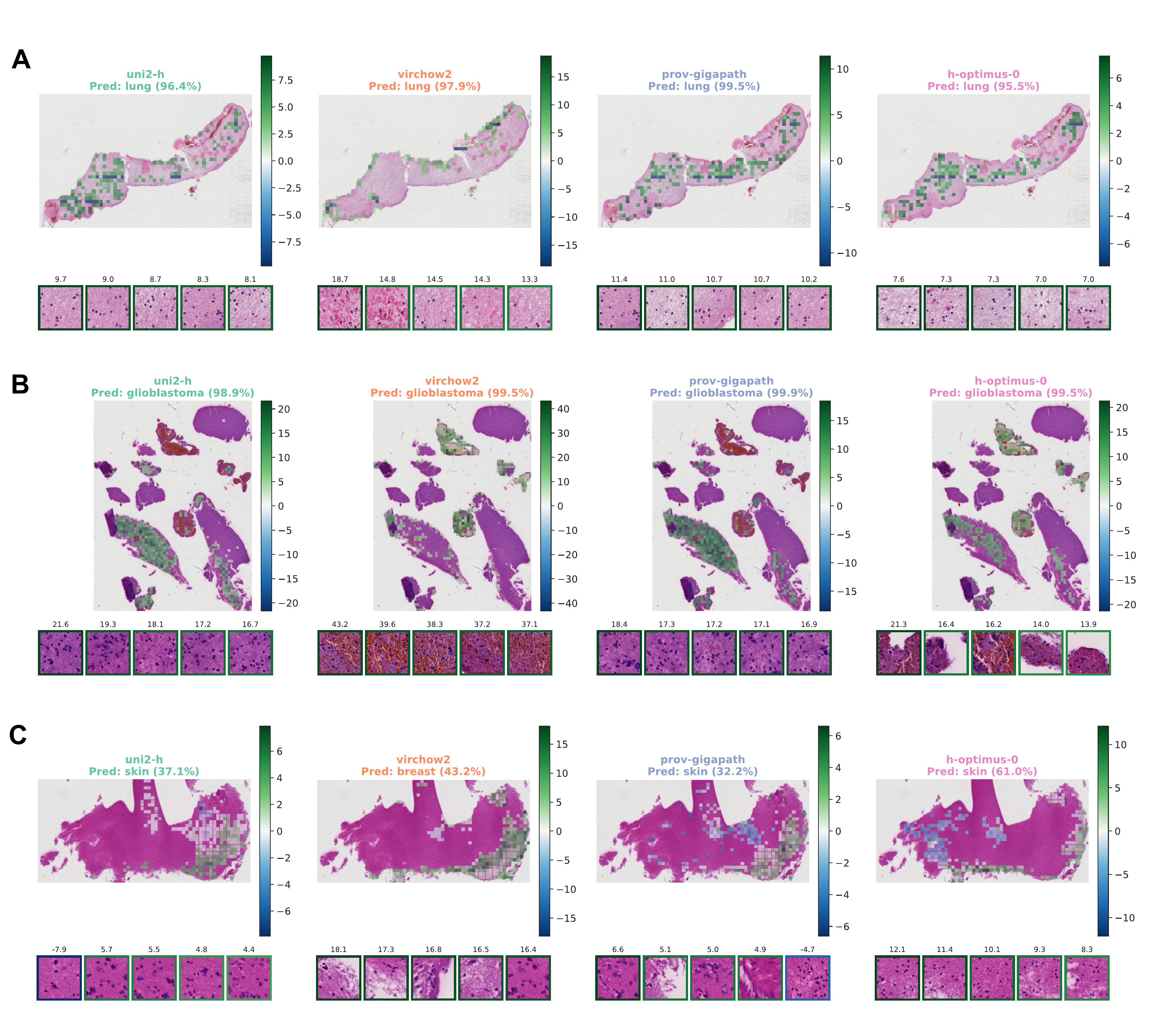}
    \caption{%
Comparative attention maps across four foundation-model encoders (uni2-h, virchow2, prov-gigapath, h-optimus-0) for three representative slides. 
In each panel, the top row overlays signed attention on the whole-slide thumbnail for every encoder, and the bottom row shows the five most influential tiles (ranked by absolute signed attention). 
The diverging colormap encodes the sign of each tile's contribution toward the predicted class (green) versus against it (blue); intensity reflects attention magnitude. 
Panel titles give each encoder's prediction and confidence, pooled across Monte~Carlo folds. 
\textbf{(A)} Metastatic-origin, a lung metastasis correctly classified.
Attention concentrates on reactive, non-lesional parenchyma rather than lesional tissue, consistent with the ground-truth curation.
\textbf{(B)} Glioma-typing, a glioblastoma correctly classified with high confidence by all encoders (consensus-correct). 
\textbf{(C)} Metastatic-origin, a true breast metastasis for which the consensus pools to an incorrect skin prediction, the dominant failure mode for this task.%
}
    \label{fig:attention}
\end{figure*}

All results are reported as the mean $\pm$ standard deviation over the respective Monte Carlo cross-validation folds, with each model selected at last epoch (\autoref{tab:results}).

\paragraph{Slide size is a confound for the coarse category task.}
Because our reactive-tissue labels are obtained by two sampling strategies that yield systematically different slide sizes (\autoref{sec:methods}), we tested whether the label is recoverable from slide size alone.
A logistic-regression classifier trained only on the log tile count per slide, using the same folds and metrics, reaches a balanced accuracy of $0.782$ (AUC $0.877$) on the category task, well above the $0.33$ chance level and close to the best encoder ($0.798$).
Slide size therefore accounts for a large fraction of the above-chance category signal, and the category results should be read as partially size-driven rather than as evidence of a morphological tissue response.
In contrast, the size-only baseline is near chance for the two finer-grained tasks metastatic origin ($0.186$ balanced accuracy; AUC $0.543$) and glioma typing ($0.361$; AUC $0.481$) and only marginally above chance for tumor subtyping ($0.518$; AUC $0.535$).
The morphological signal that the foundation-model encoders recover on these tasks is thus not explained by slide size.

\paragraph{All encoders learn a discriminative signal.}
On the two coarse-grained tasks, every encoder performed well above chance.
Broad tissue-category classification (chance $0.33$) reached a BACC of $0.798$--$0.792$ and an AUC of $0.921$--$0.912$, and glioma-versus-metastasis subtyping (chance $0.50$) reached a BACC of $0.807$--$0.819$ and an AUC of $0.906$--$0.921$.
The tight spread across encoders and the high AUC values indicate that the aggregated whole-slide representations carry a strong, consistently recoverable signal for these distinctions, largely independent of the choice of foundation model.

\paragraph{Performance degrades with task granularity.}
Accuracy decreased as the tasks moved deeper into the diagnostic hierarchy.
Glioma typing (chance $0.33$) yielded a BACC of $0.437$--$0.581$ (AUC $0.655$--$0.747$), while classifying the metastatic origin (four classes, chance $0.25$) was the hardest task, with a BACC of $0.371$--$0.416$ and an AUC of $0.652$--$0.697$.
The metastatic-origin task also exhibited among the largest fold-to-fold variance ($\pm0.117$--$0.166$ BACC), consistent with the finer-grained distinction and the smaller number of samples per origin class.
Although modest, performance remained above chance across all encoders, indicating that even these subtle distinctions leave a detectable morphological trace.

\begin{figure}
    \centering
    \includegraphics[width=\linewidth]{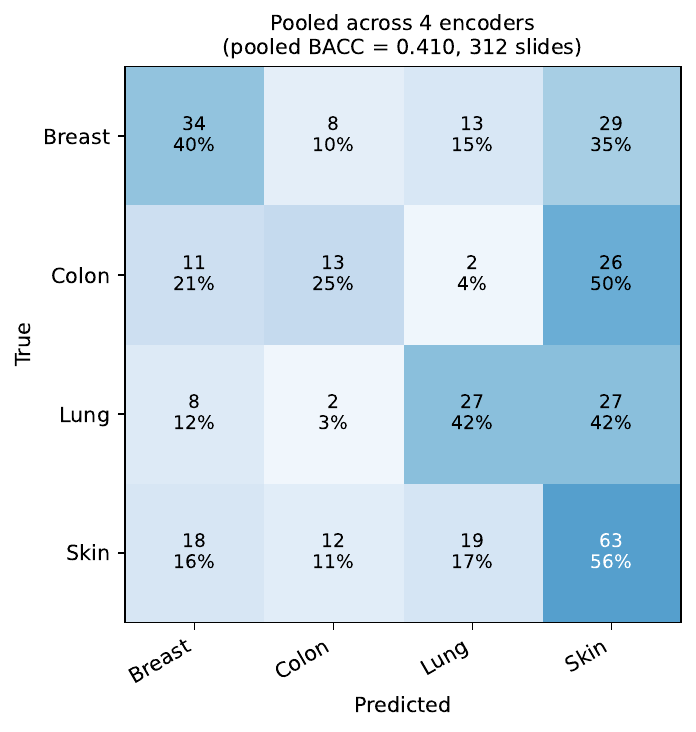}
    \caption{%
Metastatic-origin confusion, pooled across the four encoders.
Rows are true origin, columns predicted; cells give prediction counts with row-normalised percentages.
Each encoder's per-slide prediction is the argmax of its cross-validation-averaged class probabilities; 
the four encoders' predictions are stacked rather than ensembled: each of 78 slides (\autoref{sec:methods}) contributes four predictions, $N=312$. 
Balanced accuracy $=0.410$ (pooled).
}
    \label{fig:confusion}
\end{figure}

\paragraph{Attention localises to reactive, non-lesional tissue.}
Attention maps for representative slides are shown in \autoref{fig:attention}.
Across the correctly classified cases (\autoref{fig:attention}A,B), the signal supporting each diagnosis is recoverable from the reactive brain tissue that does not itself contain the lesion. 

For the consensus-correct lung metastasis (\autoref{fig:attention}A), all four encoders agree and localise to similar regions, indicating that the discriminative signal is not an artifact of a single model.
Similarly, in the correctly classified glioblastoma case (\autoref{fig:attention}B) the encoders broadly agree on prediction while attending to different parts of the image.

The failure case (\autoref{fig:attention}C) is instructive: 
a true breast metastasis that the consensus pools to a skin prediction, the dominant failure mode for the metastatic-origin task.
Three of the four encoders (uni2-h, prov-gigapath, h-optimus-0) converge on morphologically similar tissue regions and all pool to skin, while virchow2 attends to a distinct set of tiles and is the only encoder to recover the correct breast origin, albeit with low confidence ($43.2\%$); this partial disagreement between encoders is consistent with the generally low, borderline pooled probabilities across all four models ($32.2-61.0\%$) rather than one model being clearly correct and the others clearly wrong.

This mirrors the aggregate error structure (\autoref{fig:confusion}): the above-chance origin performance (macro AUC $0.652 - 0.697 $ / balanced accuracy $0.371 - 0.416$, \autoref{tab:results}) is not carried by a single easy class, with the residual confusion dominated by over-prediction of skin (every non-skin origin misclassified as skin in $35 - 50\%$ of slides, per-encoder breakdowns in Supp.~S3).
As skin is the largest metastasis class, this bias is at least partly a prevalence effect.

\paragraph{Blood as a confounder in category prediction}
Attention maps also revealed a sampling confounder.
A subset of the top-ranked attention tiles captured blood introduced during tissue sampling rather than disease-relevant tissue, and the frequency of such tiles did not differ between correctly and incorrectly classified cases, indicating that blood attention does not by itself drive the category predictions.
We quantified its prevalence on the category task, chosen because it comprises the most slides and therefore affords the highest statistical power for a per-tile analysis ($183$ slides analysed).
Across the four encoders the proportion of blood-containing patches among the top-5 attention tiles varied considerably: $27.0\%$ for Virchow2, $15.7\%$ for H-optimus-0, $4.4\%$ for Prov-GigaPath, and $4.0\%$ for UNI2-h, a nearly sevenfold spread (Friedman test across the four encoders, $\chi^2(3) = 151.6$, $p < 0.001$).

The same tendency was visible qualitatively on the finer task (\autoref{fig:attention}B), where Virchow2 (all five selected tiles) and H-optimus-0 (tiles~1, 3, and~5) again foregrounded patches containing substantial erythrocyte content.
These findings indicate that some foundation models are markedly more susceptible than others to attending to sampling artifacts unrelated to the underlying pathology.

\paragraph{Encoders separate from chance but not from each other.}
Every encoder exceeded chance on every task for both metrics: 
the conditional permutation test reached its resolution floor in all $32$ cases ($4$ tasks $\times$ $4$ encoders $\times$ $2$ metrics; $p \le 1\times10^{-4}$, no permutation out of $10{,}000$ matched or exceeded the observed score), including the hardest metastatic-origin task.
In contrast, no pairwise difference between encoders survived correction. 

Across all four tasks and both metrics ($48$ comparisons in total), no pairwise difference was significant after Holm correction (smallest $p_{\text{Holm}} = 0.25$). 
The smallest uncorrected $p$-value was $0.042$, which does not survive correction for multiple testing.
The apparent per-task ``winners'' in \autoref{tab:results} therefore fall within fold-to-fold noise.
We retain boldface only to mark the numerically highest value, not a significant one.
The practical implication is that, for this problem, the choice of pathology foundation model has no detectable effect on performance:
what matters is that a strong general-purpose encoder is used, not which one.
We note that with $11$--$20$ folds we are underpowered to resolve differences of the observed magnitude, so this is evidence of equivalence at that resolution rather than a claim that the encoders are identical.

\section{Discussion}
\label{sec:discussion}

We framed the identification of the disease process underlying reactive gliosis as a weakly supervised whole-slide classification problem, and asked whether pathology foundation models can recover disease-specific signal from tissue that lacks overt lesional morphology.
Two findings structure our interpretation.
First, coarse tissue-category classification is substantially confounded by slide size:
a logistic-regression baseline on log tile count alone comes within roughly two balanced-accuracy points of the best encoder, so this task cannot be read as evidence of a learned morphological tissue response.
Second, and in contrast, the two finer-grained tasks (glioma typing and metastatic-origin classification) are not recoverable from slide size.
The size-only baseline sits at or near chance on both, yet the encoders classify them above chance under label-permutation testing.

\paragraph{The size confound and what it does and does not undermine.}
The two ground-truth strategies necessarily produce slides of different extent, and we deliberately measured rather than assumed the consequence.
The size-only baseline is above chance only on the category task and falls to chance on every finer distinction, which localizes the confound cleanly and is why we report it as a standard control rather than a caveat.
Beyond our cohort, this illustrates a general risk: 
whenever label provenance correlates with a low-level image statistic such as tissue extent, a model can exploit it, and image-only benchmarks will not reveal the shortcut. 
Reporting a provenance-only baseline alongside image models is a cheap, general check.
We note, however, that the baseline detects size as a \emph{correlate} of sampling provenance; it cannot separate tile count \emph{per se} from tissue- or batch-level differences that travel with provenance, such as staining or scanning variation between excised margins and whole biopsies.
A fixed-tile-budget analysis would remove count as a feature but not such covariates; a provenance-stratified evaluation is the more complete control and a priority for future work.

\paragraph{Equivalent accuracy can mask divergent attention.}
Although the encoders are statistically indistinguishable at the slide level, their attention is not: the fraction of top-ranked tiles capturing blood introduced during sampling rather than tissue ranged from $4.0\%$ (uni2-h) to $27.0\%$ (virchow2), a nearly sevenfold spread. 
Equivalent performance therefore does not imply equivalent behavior.
Models reach similar predictions while weighting the slide differently, and some are markedly more prone to sampling artifacts (Supp.~S3).
Because this artifact rate did not differ between correct and incorrect cases, it does not by itself explain errors, but it is a caution for interpretability:
attention maps from a single encoder may foreground preparation artifacts, and the choice of model for \emph{explanation} carries considerations that its choice for \emph{prediction}, by our results, does not.

\paragraph{Single-site cohort and external validity.}
All slides originate from a single institution, and the cohort is modest in size, particularly for the metastatic-origin task, where the per-class support is small, only a subset of folds could be evaluated, and the fold-to-fold variance is correspondingly large.
We therefore make no claim about cross-site generalization: the reactive-tissue signal we report could in principle reflect institution-specific staining or scanning characteristics rather than transferable biology.
For the same reason we emphasize the \emph{existence} of the signal, established by permutation testing, over the precise magnitude of any single score, whose confidence intervals on the finest tasks are wide (Supp.~S1).
Establishing that these signatures persist across scanners, protocols, and centers is the essential next step before any diagnostic interpretation, and we present these results as evidence of feasibility on a well-characterized cohort rather than as a validated classifier.

\paragraph{Representation choice is not the bottleneck.}
Across all four tasks and both metrics, no pairwise difference between encoders was statistically significant after correction for the overlap between cross-validation folds, despite every encoder separating clearly from chance.
The apparent per-task ordering in \autoref{tab:results} therefore falls within fold-to-fold noise, and we caution against reading it as a ranking of foundation-model quality.
That four models with markedly different pretraining data, objectives, architectures, and receptive fields converge to indistinguishable performance suggests that, on reactive tissue at the patch level, the limiting factor is not the choice of representation but the intrinsic subtlety of the signal and the scale of the cohort.
We read this as a pointer rather than a dead end: because all encoders share a common $256$-pixel patch tokenization, the plateau may reflect that granularity averaging over nuclear-scale morphology, motivating representations operating below the patch level for example at the resolution of individual nuclei as a direction for recovering signal that patch-level encoders leave on the table.

\paragraph{Linking morphology back to biology.}
The subtle, spatially diffuse signal recovered here is consistent with the view that reactive gliosis is not a uniform response but comprises molecularly distinct, context-dependent astrocytic and microglial states \cite{Hasel2021-pa,Keren-Shaul2017-we}. The signed contribution maps localize where this signal resides but do not resolve which cellular programs underlie it. Pairing this weakly supervised histology framework with single-cell or spatial transcriptomic profiling of the same reactive regions could help determine whether the tile-level signatures that drive correct classification correspond to specific reactive glial states, offering a route toward a mechanistic, rather than purely predictive, understanding of how the tissue's response to a lesion encodes information about its cause.

\ifreview\else
\section*{Acknowledgements}
J.E. was supported by the Medical Scientist Kolleg InFlame funded by the Else Kröner-Fresenius Foundation, Germany.
\fi

{
    \small
    \bibliographystyle{ieeenat_fullname}
    \bibliography{main}
}

\end{document}